\documentclass[transmag]{IEEEtran}
\usepackage{latexsym}
\usepackage{graphicx}
\graphicspath{ {./Figures/} }
\usepackage{xcolor}
\usepackage{dsfont}
\usepackage[noadjust]{cite}
\usepackage{amsfonts,amssymb,amsmath}
\usepackage[ruled,vlined]{algorithm2e}
\usepackage[noend]{algpseudocode}
\usepackage{hyperref}
\def\BibTeX{{\rm B\kern-.05em{\sc i\kern-.025em b}\kern-.08em T\kern-.1667em\lower.7ex\hbox{E}\kern-.125emX}}
\DeclareMathOperator*{\argmax}{arg\,max}
\DeclareMathOperator*{\argmin}{arg\,min}

\DeclareMathOperator*{\clip}{clip}
\begin{document}

% \title{Robust and Scalable Routing with Multi-Agent Deep Reinforcement Learning for Dynamic Mobile Ad-Hoc Networks}
\title{Robust and Scalable Routing with Multi-Agent Deep Reinforcement Learning for MANETs}
% \author{Saeed Kaviani, Bo Ryu, Ejaz Ahmed, Kevin Larson, Anh Le, Alex Yahja, and Jae H. Kim
\author{\IEEEauthorblockN{Saeed Kaviani\IEEEauthorrefmark{1}, Bo Ryu\IEEEauthorrefmark{1}, Ejaz Ahmed\IEEEauthorrefmark{1}, Kevin Larson\IEEEauthorrefmark{2}, Anh Le\IEEEauthorrefmark{1}, Alex Yahja\IEEEauthorrefmark{1}, and Jae H. Kim\IEEEauthorrefmark{2}
    \\\IEEEauthorblockA{\IEEEauthorrefmark{1}EpiSys Science, Inc. 
    \\\{saeed, bo.ryu, ejaz, anhle, alex\}@episci.com} \\
    \IEEEauthorblockA{\IEEEauthorrefmark{2}Boeing Research and Technology
    \\\{kevin.a.larson, jae.h.kim\}@boeing.com}}

% \thanks{S. Kaviani, B. Ryu, E. Ahmed, A. Le, and A. Yahja are with the EpiSys Science, Inc. K. Larson and J. H. Kim are with the Boeing Research and Technology.}
}

\maketitle

\begin{abstract}
Highly dynamic mobile ad-hoc networks (MANETs)  are continuing to serve as one of the most challenging environments to develop and deploy robust, efficient, and scalable routing protocols. In this paper, we present \textbf{DeepCQ+ routing} which, in a novel manner, integrates emerging multi-agent deep reinforcement learning (MADRL) techniques into existing Q-learning-based routing protocols and their variants, and achieves persistently higher performance across a wide range of MANET configurations while training only on a limited range of network parameters and conditions. Quantitatively, \textbf{DeepCQ+} shows consistently higher end-to-end throughput with lower overhead compared to its Q-learning-based counterparts with the overall gain of 10-15\% in its efficiency. Qualitatively and more significantly, \textbf{DeepCQ+} maintains remarkably similar performance gains under many scenarios that it was \emph{not} trained for in terms of network sizes, mobility conditions, and traffic dynamics. To the best of our knowledge, this is the first successful demonstration of MADRL for the MANET routing problem that achieves and maintains a high degree of scalability and robustness even in the environments that are \emph{outside} the trained range of scenarios. This implies that the proposed hybrid design approach of \textbf{DeepCQ+} that combines MADRL and  Q-learning significantly increases its practicality and explainability because the real-world MANET environment will likely vary outside the trained range of MANET scenarios. 

\end{abstract}

\section{Introduction}

% Routing overview (distributed)
Routing has been one of the most challenging problems in communication and computer networks, especially in uncoordinated distributed and autonomous wireless networks. In packet routing protocols, each router selects another node as a next-hop (i.e. unicasting) as determined by its routing policy. The level and type of cooperation and coordination between the nodes in the network often describe what information the routing decisions are based on. Distributed routing algorithm can assign a designated communication channel or messaging architecture to accommodate the cooperation. We concentrate on the algorithms that only share limited information through acknowledgment (ACK) packets. This information sharing is efficient as ACKs are inherently present in the networking protocols and do not require any extra implementation in the system.  

% Routing as a next-hop selection 
Our work is focused solely on optimizing next-hop selection and mode (broadcast or unicast) selection, with the primary goal of achieving low overhead in environments with highly dynamic networks. Additional performance is likely available through network coding (NC) methods (such as those based on packet manipulation and/or designated coordination channels), but these are outside the scope of this work, and not considered. \cite{ahlswede2000network,biswas2005exor,chachulski2007trading,katti2008xors, perin2020maximizing}. A common theme in many of these algorithms is opportunistic routing (OR). OR has many disadvantages, such as: 1) Over-reliance on costly broadcast and coordination traffic. Unicast communication is sufficient in most cases. 2) Broadcast is incompatible with the directional transmission, and will result in lower data rates and higher end-to-end delay in wireless channels with beamforming capabilities when compared with unicast transmissions. 3) OR requires designated coordination channels between the nodes receiving broadcasted packets. For example, in the ExOR multi-hop opportunistic routing algorithm \cite{biswas2005exor}, all nodes broadcast the packet.  Then, each node that receives the packet sends additional traffic to reach a consensus on which node will broadcast (forward) that packet next. ExOR's designated coordination channel and packet broadcasting add considerably to the communication overhead. 4) Due to these coordination steps, these routing protocols are too slow to be robust and reliable for highly dynamic MANET. Opportunistic multi-hop routing methods like MORE \cite{chachulski2007trading}, COPE \cite{katti2008xors}, and recently  \cite{perin2020maximizing}, all rely on NC (mixing the packets) to extend the useful context in forwarded traffic. The gains in these methods are based on packet manipulation and high-cost coordination messaging rather than simply leveraging context in existing traffic, such as ACK messages. Nevertheless, our proposed approach considers routing exclusively from the context of next-hop selection and offers an optimization framework that could be extended by NC for additional improvement.

% MANET and tactical networks routing needs -> robust and reliable
The challenges of routing are compounded when the network is highly dynamic, heterogeneous, and/or contains variable data rates. These challenges are of particular for MANET in the tactical networking domain. The term "Tactical network" is used for the wireless network supporting tactical operations, where reliable wireless networking is required for critical operations, including those of military, broadcast, unmanned vehicles, law enforcement, and airborne intelligent surveillance reconnaissance (ISR). Tactical wireless communications have latency, reliability, and security requirements not found in commercial wireless networks and pose many challenging problems in the routing domain. The nodes in tactical networks are often highly mobile, expressing unpredictable motion and topology  due to environmental conditions such as terrain, interference, and jamming \cite{elmasry2010comparative,pawgasame2015tactical,taneja2010survey}. These factors significantly increase uncertainty in tactical networks in hostile environments. As a result of these environments, many traditional MANET routing protocols are unreliable and require re-computation of end-to-end routes on every network change, resulting in a periodic loss in throughput due to traffic not being sent during routing table computations. To improve packet delivery and rapid exploration in these highly dynamic networks, broadcasting (i.e. transmission of a packet to all neighbors) has become a popular technique \cite{taneja2010survey,perkins1994highly, perkins1999ad, clausen2003optimized, moy1998ospf}. Traditional network routing protocols (e.g. DSDV \cite{perkins1994highly}, OSPF \cite{moy1998ospf}, OLSR \cite{clausen2003optimized},  and AODV \cite{perkins1999ad}) are used when the network is in a stable state.  When link outages and node mobility become too frequent, these algorithms require alternative strategies to sustain performance. Danilov et al. \cite{danilov12}  discussed the poor performance of these link-state routing protocols in tactical environments and attempted to reduce loss during transitions by flooding.

Although it seems that the wireless channel broadcasts all transmissions (so any node in range could receive the transmitted data packet), packets marked for unicast and broadcast are treated differently. When broadcasting, all that receive the packet process it as an intended transmission. On the other hand, this is not the case for unicast, as nodes not marked as the destination simply drop the packet. When directional antennas are available, unicast can use a single beam, where broadcast requires omnidirectional antennas. Therefore, unicast produces higher received signal power (higher signal-to-noise-ratio (SNR)) and can deliver packets at a higher data rate (reduce over the air time, etc.)

%Robustness
To meet the demands of highly dynamic MANET, many solutions have adapted routing protocols to variations in the network conditions, e.g. fish-eye state routing protocol (FSR) \cite{pei2000fisheye} use adaptive link-state update rates, and the adaptive distance vector (ADV) routing protocol \cite{boppana2001adaptive} use a threshold-based adjustment of the routing update rates based on the network dynamics. 
While these protocols outperform traditional routing schemes with lower overhead and topology information sharing, they are not responsive enough in highly dynamic MANETs where the routing recalculations happens at slower paste than link changes. 

% RL-based routing: Q-routing
 The seminal work in \cite{Boyan1994} proposed Q-routing, which uses a reinforcement learning (RL) module (i.e. Q-Learning \cite{sutton2018reinforcement}) to route packets and minimize delivery time. Each node uses $Q$-values based on locally acquired statistics to determine the next hop. Each $Q$-value represents the quality of each next-hop (or route) as an estimation of the delay for each path. The $Q$ data is shared only via ACK messages, and each node maintains a table of values for every neighbor and destination pair. After any transmission, a node may receive an ACK message containing values to update the $Q$ table. The Q-routing protocol selects the next hop with the best $Q$-value. Q-routing is efficient in static and minimally dynamic networks. In dynamic networks, $Q$-values quickly become stale as links break.
 
    % Q-routing + adaptive updates: CQ-routing
Kumar et al. \cite{kumar98} improved Q-routing for dynamic networks with the addition of confidence values (i.e. $C$-values) in their \textit{CQ-routing protocol}. $C$-values are incremented when $Q$-value is updated and decremented as it becomes stale; however, CQ-routing becomes inefficient in highly dynamic networks as it is based on only uni-casting to a single node. With only unicast transmissions, network exploration (Q-value updates) is too slow when the network changes rapidly. The rate at which information is shared is simply too slow to keep up with highly dynamic networks.

% CQ-routing + broadcast (R2DN) CQ+-routing
 AR \cite{danilov12} and smart robust routing (SRR) algorithms \cite{johnston18} have attempted to supplement unicast transmission with broadcast to improve robustness in highly dynamic networks.  Both of these algorithms revert to unicast transmission in order to reduce the overhead of the flooding. Johnston et al \cite{johnston18} use techniques from the CQ-routing protocol (i.e. $C$ and $Q$-values) but extends it by adding broadcast procedure for high reliability, robustness, and rapid network exploration as needed in the tactical and highly dynamic MANETs. To simplify the convention when listed with the other protocols, we will refer to it as the CQ+ routing protocol. 
%  For example, when the network parameters (i.e. $C$ and $Q$ values) are outdated or not available, it is preferred to perform broadcast as the transmitter may receive multiple ACKs from neighbors to immediately explore its surroundings and/or to have a lower probability of failure. 
 Although CQ+ routing uses a simple but efficient switching policy to choose between unicast and broadcast, its decisions depend on a single network parameter (best path confidence level). It has a limited perspective of the entire network and can settle on a locally optimal solution. CQ+ routing also does not account for the change rate of network parameters and congestion in forwarding paths. We build more perspective into traffic and queuing and leverage this information to further improve performance. Among the various routing algorithms for MANET networks, Q-routing, CQ-routing, and CQ+ routing approaches are considered as benchmarks for this work. 
 
% RL based approaches
Routing decisions, such as next-hop selection, are opportune targets of reinforcement learning (RL). The work in this area was initiated by Boyan's Q-routing protocol\cite{Boyan1994}. Following the Q-routing approach, many other techniques and algorithms from the RL community have been applied to packet routing and scheduling \cite{you2020toward, ali2020hierarchical, mammeri2019reinforcement, yu2018drom, stampa2017deep, ye2019deep, lin2016qos, valadarsky2017}. \cite{you2020toward} uses MADRL to design independent deep routing policies for each agent based on an off-policy deep Q-learning RL algorithm. Deep Q-learning approaches are based on value estimation, an estimation of the expected reward of the actions at certain states. Consequently, deep Q-learning and value estimation policies scale poorly, as the expected reward is dependent on many network parameters and conditions that are not known prior to decisions. Moreover, in MADRL-based approaches in the literature like \cite{you2020toward}, training unique policies for each agent further limits scalability. It is unclear how policies trained for specific network sizes perform when the network is extended or shrunk. A similar deep Q-learning RL-based approach is used in \cite{ali2020hierarchical}, where it creates cluster abstractions in the network and accounts for inter-cluster routing performance. They also assume a feedback link is available from the source node to the cluster lead agents. Although it is claimed that their approach is expected to be scalable to larger networks and dynamics but it is not clear how this would scale and perform if only trained on smaller networks and limited network settings. Deep neural network (DNN)-based routing policies tend to struggle with high dynamics, as the complexity of multi-agent environment struggles with the rapid rate of change. Optimization of the policy for one agent is dependent on the policy and actions of other agents and therefore suffers from non-stationarity. This is particularly difficult in dynamic networks as many network parameters and topology are rapidly changing. 

To the best of our knowledge, there have not been works on a scalable and robust routing policy design framework using MADRL in MANET. To provide robustness and reliability, we use a similar approach to CQ+ routing, where CQ-routing is combined with adaptive flooding. In this paper, we use MADRL and advanced RL algorithms and techniques to train a robust and reliable routing policy that can be applied to any network size, traffic, and dynamic. More importantly, the MADRL-based framework includes techniques and formulations that enable us to train on a limited range of network parameters (e.g. smaller network sizes, single data flow, small variations in dynamic level, average node velocities, and network coverage area size) but test and execute the trained policies on a wider range of configurations. This is significant as due to the curse of dimensionality training over large network sizes and wide range of configurations are not practical. Our proposed framework and designed policies are closely related to the CQ+ routing protocol, therefore we refer to it as deep robust routing for dynamic networks (\textbf{DeepCQ+ routing}).

\section{Robust Routing Framework}
We consider a robust routing protocol that monitors the quality and confidence level of the routes (next-hops) via ACK messages, i.e. CQ-routing protocol \cite{kumar98}. We use this protocol as it does not add any designated communication link for coordination between nodes in the network, and it is robust for dynamic networks with the addition of confidence level, i.e. $C$. We specifically use the extended version of CQ-routing by addition of adaptive broadcasting to bring more reliability (higher delivery rates) and rapid network exploration (for highly dynamic networks), as introduced in \cite{johnston18} and we refer to it as CQ+ routing. We summarize the CQ+  routing algorithm in Algorithm \ref{SRR}, which shows how it chooses between broadcast and unicast (and next-hop) adaptively. 

\subsection{SRR (CQ+ routing) Protocol}\label{section:r2dn_protocol}
The smart robust routing (SRR) algorithm proposed in \cite{johnston18} uses the network parameters $C$ and $H$\footnote{We have renamed the original $Q$-factor in CQ-routing to $H$-factor to prevent confusion with the $Q$ in the $Q$-networks and/or $Q$-learning in the RL context.} for the routing decisions primarily introduced in seminal CQ-routing \cite{kumar98}. Each node $i$ has a $H$-factor, $h(i,j,d)$ (i.e. $i \rightsquigarrow j \rightsquigarrow d$), which represents an estimate of the least number of hops between node $i$ and destination $d$ which passes through potential next-hop $j$. To monitor the dynamics of the network, each node $i$ also have a confidence level or $C$-value, $c(i,j,d)$, that represents the confidence in likelihood the packet will reach its destination $h(i,j,d)$. This $C$-value is increased and corrected with any packet transmission success (receiving the ACK). Every packet transmission, the $C$-value is degraded by a decay factor.

$C$- and $H$-factors are updated through the $c_\text{ack}$ and $h_\text{ack}$ which are propagated by the acknowledgement (ACK) packets from the receiving node (e.g. next-hop) to the transmitting node. These ACK values are computed at the next-hop node $j$ as 
\begin{equation}\label{h_ack}
h_{\text{ack}} = 1 + h(j,\hat{k},d)
\end{equation}
\begin{equation}\label{c_ack}
c_{\text{ack}} = c(j,\hat{k},d)
\end{equation}
where $\hat{k}$ is the best path (next-hop) estimate of node $j$ to destination $d$ and it is found by
\begin{equation}
\hat{k} = \argmin\limits_k h(j,d,k)\left(1-c(j,k,d)\right)    
\end{equation}
In other words, $C$- and $H$-values are exponential moving average of the $c_\text{ack}$ and $h_\text{ack}$, respectively. If a transmission fails or there is no ACK to update $C$- and $H$-levels, then $H$-level cannot be updated. However, we degrade the $C$-level as in $c_\text{ack} = 0$ to reflect the path failure. The updates of the $C$- and $H$-levels given by 
\begin{equation}\label{update_H}
    h_{t+1}(i,j,d) = (1-\alpha)h_t(i,j,d) + \alpha h_\text{ack},
\end{equation}
\begin{equation}\label{update_C}
    c_{t+1}(i,j,d) =
\begin{cases}
 (1-\lambda)c_t(i,j,d) & \text{failure}\\
 (1-\lambda)c_t(i,j,d) + \lambda c_\text{ack} & \text{otherwise} 
\end{cases}
\end{equation}
where $0 \leq \alpha \leq 1$ is a discount factor for the new observation with an adaptable value of 
\begin{equation}\label{eqn: discount_factor_alpha}
\alpha = \max\left(c_\text{ack}, 1- c_t(i,j,d)\right).
\end{equation}
$\lambda$ is the decay factor for the new observation of 
$c_\text{ack}$ (or $1-\lambda$ is the decay factor for the old observation). If packet is received at the destination $d$, then $c_\text{ack}$, and $h_\text{ack}$ are set to 1, indicating that we have full confidence that we are 1 hop away from destination.
The SRR algorithm (CQ+ routing) is summarized in Algorithm \ref{SRR}, where more details are available in \cite{johnston18}. SRR algorithm includes \textit{(i)} reception of ACK and consequently updating $C$- and $H$-levels, \textit{(ii)} reception of non-ACK packets by checking for duplication, loops, and pushing into the queue, \textit{(iii)} transmission of data packets from the queue by routing it according to the policy in use. We simply refer to this decision policy as CQ$^+$- routing policy.   
\begin{algorithm}
\SetAlgoLined
\caption{SRR (CQ+) algorithm \cite{johnston18}}
\label{SRR}
\small
{Receive incoming packet at node $i$:} \\
\eIf{Packet is ACK}
    {Update $c$ and $h$ from (\ref{update_H}) and (\ref{update_C})\;}
    {\If{packet traversed a loop}
        {Drop packet, do not return ACK\;}
     \If{packet is already in queue}
        {Drop packet \\
        Find best next-hop $j^\star$ from (\ref{best_next_hop}) \\
        Compute $c_\text{ack}$ and $h_\text{ack}$ from (\ref{c_ack}) and (\ref{h_ack}) using $j^\star$ \\
        Return ACK}
     \eIf{packet is not duplicate}
        {Add packet to the queue}
     {Do not add packet to the queue}
    }
\If{Queue is not empty}
    {Pick up packet from queue \\
    \mbox{\textsc{Routing Decision Policy}} \\
    $P_\text{BC} \leftarrow \epsilon + (1-\epsilon)(1-c(i,j^\star,d))$ \\
    Choose $\begin{cases}
     \text{Broadcast} & \text{with probability } P_\text{BC} \\
     \text{Unicast to } j^\star & \text{with probability } 1 - P_\text{BC}
     \end{cases}$ \\
     \If{Broadcast}{Forward packet to all}
     \If{Unicast}{Forward packet to $j^\star$}}
\end{algorithm}

\subsubsection{SRR (CQ+ routing) policy}
The SRR policy chooses the next-hop based on the minimization of the information uncertainty and the expected number of hops. This is given by
\begin{equation}\label{best_next_hop}
    j^\star = \argmin_j h(i,j,d)\left(1-c(i,j,d)\right).
\end{equation}
The next-hop $j^\star$ is considered by the CQ+ routing policy if it unicast the packet to one single node. However, the CQ+ routing policy enables broadcasting to minimize the next-hop information uncertainty. The uncertainty of the next-hop is measured by $1 - c_i(d,j^\star)$. Therefore, CQ+ routing decides to broadcast if the uncertainty about information obtained is high to explore the network more and make a reliable transmission by flooding the neighbors. The CQ+ routing policy is probabilistic and it assigns the probability of broadcast as 
\begin{equation}\label{r2dn_policy}
    P_\text{BC} = \epsilon + (1-c(i, j^\star,d))(1-\epsilon) = 1 - c(i, j^\star,d)\tilde{\epsilon} 
\end{equation}
where a small value $\epsilon$ is used for the minimum probability of broadcast (defined for exploration purposes) and correspondingly $\tilde{\epsilon} = 1 - \epsilon$ is the maximum probability of unicast. 

In this paper, we use the CQ+ routing protocol, meaning that the reception procedure of our algorithm is the same as described in \ref{SRR}, but our improvements are applied to the transmission procedure and the CQ+ routing policy. In this work, we preserve the protocol and concentrate on routing policy optimization.

\section{Deep Reinforcement Learning for CQ routing (DeepCQ+)}

\subsection{Deep Reinforcement Learning Background}
Our robust routing problem fits as a \textit{decentralized partially-observable Markov decision process (Dec-POMDP)}, that is used for decision-making problems in a team of cooperative agents \cite{oliehoek2016concise}. Dec-POMDP tries to model and optimize the behavior of the agents while considering the environment's and other agents' uncertainties. POMDP is defined by a set of states $\mathcal{S}$ describing the possible configurations of the agent(s), a set of actions $\mathcal{A}$, and a set of observations $\mathcal{O}$. The actions are selected using a stochastic policy $\pi_\theta: \mathcal{O}\times\mathcal{A} \rightarrow [0,1]$ (parameterized by $\theta$), which results to the next state defined by the environment's state transition function $\mathcal{T} : \mathcal{S} \times \mathcal{A} \rightarrow \mathcal{S}$. As a result of this transition, a reward is obtained by an agent(s) and it is described as a function of the state and action, i.e. $r: \mathcal{S}\times \mathcal{A} \rightarrow \mathbb{R}$. Each agent\footnote{In the Dec-POMDP, these sets are defined for each agent separately} receives an observation related to the state as $o : \mathcal{S} \rightarrow \mathcal{O}$. Below, we give an overview and brief introduction of the DRL algorithms and techniques used in this work to design the DeepCQ+ routing policies. 

\subsubsection{Value Optimization and Deep Q-learning }
To find the optimal policy, Q-Learning\footnote{The name of the Q-routing algorithm is inspired by the Q-learning algorithm in RL but it only represents the expected number of hops for different paths as action-value function.} is a widely used model-free algorithm that estimates the action-value function for policy $Q^\pi(s, a)$. The action-value function (or Q-function) is the expected maximum sum of (discounted) rewards, perceived at state $s$ when taking action $a$ and it is given by 
\begin{equation}
    Q^\pi(s,a) := \mathbb{E}\left[\sum_{t=0}^T\gamma^tr_t \bigg | s_t = s, a_t = a\right]
\end{equation}
where $\gamma$ is the discount factor and $T$ is the time horizon. The action-value function can be recursively written (and calculated) as $Q^\pi(s,a) = \mathbb{E}_{s'}[r(s,a) + \gamma \mathbb{E}_{a' \sim \pi}[Q^\pi(s',a')]]$. 

The recent deep learning paradigm enables the RL algorithms to approximate the Q-function using a deep neural network, i.e. $Q(s, a) \approx Q(s, a; \theta)$, where $\theta$ is the set of neural network parameters\footnote{In the paper, when we want to refer to neural network coefficients or weights, we have used neural network parameters. Note that we use "network parameter" term when refer to the communication network parameters such as $c$, $h$, etc.}. A popular method to do this is is known as \textit{deep Q-network (DQN)} \cite{mnih2015human}. DQN learns the optimal action-value function $Q^\star$ by minimizing the loss: 
\begin{equation}
    \mathcal{L}(\theta)\! =\! \mathbb{E}_{(s,a,r,s')}\!\left[\!\left(\!Q^\star\!(s, a; \theta)\! -\! (r\! +\! \gamma \max\limits_{a'}\bar{Q}^\star\!(s', a'))\right)^2\right]
\end{equation}
where $\bar{Q}$ is a target $Q$-function with parameters periodically (or gradually) updated with the most recent parameters $\theta$ to further stabilize the learning. DQN also benefits from a large experience replay buffer $\mathcal{B}$ of experience tuples $(s,a,r,s')$. Finally, the optimal (deterministic) policy is described as 
\begin{equation}
\pi^\star(s) = \argmax_{a \in \mathcal{A}} Q(s, a; \theta^\star),
\end{equation}
when the optimal parameters $\theta^\star$ are found.

\subsubsection{Policy Optimizations}
Policy gradient methods are another popular choice that is based on the optimization of the policy directly rather than estimate the expected return. Let $\pi$ denote a stochastic policy which assigns a probability $\pi(a_t|s_t)$ for an action $a_t$ given a state $s_t$. In the policy optimization methods, the goal is to optimize the parameters of the policy, $\theta$, to maximize the expected discounted return objective function, 
\begin{equation}
    J(\theta) = \mathbb{E}_{s_0, a_0, s_1, ...}\Bigg[\underbrace{\sum_{t=0}^T\gamma^tr_t}_{R_\tau}\Bigg],
\end{equation} 
where the experience sequence (or path) $\tau$ is denoted as $\left\{s_0, a_0, s_1, ...\right\}$ with $s_0 \sim p_0(s_0)$ and $a_{t+1} \sim \pi(a_t|s_t)$, $s_{t+1} \sim \mathcal{T}(s_{t+1}|s_t, a_t)$. The discounted return for the sequence $\tau$ is expressed by $R_\tau$. $p_0$ is the distribution of the initial state $s_0$.

% A class of popular policy optimization methods, policy gradient (PG), optimizes these policy parameters by descending toward the gradient direction $\nabla_{\theta}J(\theta)$ \cite{sutton2000policy}. 
% Now, using the $Q$-function, the gradient of the policy can be written as 
% \begin{equation}
%     g = \nabla_{\theta}J(\theta) = \mathbb{E}_{s\sim p^\pi, a \sim \pi_\theta}\left[\nabla_{\theta}\log \pi_\theta(a|s)Q_\pi(s,a)\right]
% \end{equation}

%  Now, the expected discounted return is denoted by 
% \begin{equation}
% \begin{split}
%     J(\theta) = \mathbb{E}_{s_0, a_0, ...}\left[\sum_{t=0}^\infty \gamma^t r(s_t)\right]\\
%     s_0 \sim p_0(s_0), a_t \sim \pi(a_t | s_t), s_{t+1} \sim T(s_{t+1}|s_t, a_t)
% \end{split}
% \end{equation}
 
The action-value function $Q_\pi$, and the value function $V_\pi$, and the advantage function $A_\pi$ are defined as  
\begin{equation}
Q_\pi\left(s_t, a_t\right) = \mathbb{E}_{s_{t+1},a_{t+1}, \ldots}\left[\sum_{l= 0}^T\gamma^l r_{t+l}\right],
\end{equation}
\begin{equation}
    V_\pi(s_t) = \mathbb{E}_{a_t, s_{t+1}, \ldots}\left[\sum_{l= 0}^T\gamma^l r_{t+l}\right],
\end{equation}
\begin{equation}
    A_\pi(s,a)  = Q_\pi(s,a) - V_\pi(s),
\end{equation}
where $a_t \sim \pi(a_t|s_t)$ and  $s_{t+1} \sim \mathcal{T}(s_{t+1}|s_t,a_t)$.

% \begin{equation}
%     J(\theta) = \mathbb{E}_\tau\left[R_\tau\right] = \sum_\tau P(\tau; \theta) R_\tau
% \end{equation}
A class of popular policy optimization methods, policy gradient (PG), optimizes the policy parameters $\theta$ by descending toward the gradient direction $\nabla_{\theta}J(\theta)$ \cite{sutton2000policy}. 
Now, from the gradient estimator of the Vanilla policy gradient algorithm \cite{williams1992simple}, we know that the gradient expression can be simplified as 
\begin{equation}
    \nabla_\theta J(\theta) = \mathbb{E}_t\left[\nabla_\theta \log \pi_\theta(a_t|s_t)A_t\right]
\end{equation}
Let $\eta_{\theta_t}$ denote the probability ratio 
\begin{equation}
    \eta_{\theta_t} = \frac{\pi_\theta\left(a_t|o_t\right)}{\pi_{\theta_{\text{old}}}\left(a_t|o_t\right)}
\end{equation}
and therefore $\eta_{\theta_\text{old}} = 1$.
A popular state-of-the-art policy gradient algorithm is called proximal policy optimization (PPO) \cite{schulman2017proximal, schulman2015trust}, which uses the clipped objective function 
\begin{equation}
\mathcal{L}^{\clip}(\theta) = \mathbb{E}_t\!\left[\min\left(\eta_{\theta_t}\hat{A}_t, \clip\!\left(\eta_{\theta_t},1-\epsilon, 1+\epsilon\right)\!\hat{A}_t\right)\right] 
\end{equation}
as an optimization problem of interest. PPO is indeed a family of policy optimization methods that uses multiple epochs of stochastic gradient ascent to perform each policy updates. PPO inherits the stability and reliability of the trust region methods \cite{schulman2015trust} but implemented in a much simpler way. We have used PPO algorithm for the optimization of the routing policy in the following sections. 

% \subsubsection{Overview of Actor-Critic Techniques}
% In general, there can be some states where the outcome is the same regardless of the action the agent could take; therefore, it is not always necessary to determine the state-action value at a given state $s$, $Q(s, a; \theta)$, for every action. For instance, when playing a video game consisting of moving left or right to avoid objects, trying to decide whether the optimal action is to move left or right is useless if there is no threatening object in sight. Another example is when a UE is located at the same distance from two BSs that can provide it with the same throughput. In that case, there is not a single optimal action as the result will be the same whatever BS is selected Based on this intuition, Wang et al. have introduced the notion of dueling network where $Q(s, a; \theta)$ is decomposed into a state value $V (s; \theta) = \mathbb{E}[Q(s, a; \theta)]$ and the advantage of the corresponding action $A(s, a; \theta)$ as $Q(s, a; \theta) = V (s; \theta) + A(s, a; \theta)$ \cite{wang2016dueling}. 
% \newline

\subsubsection{Centralized Training, Decentralized Execution}
In multi-agent reinforcement learning, each agent can have its policy while sharing the environment with other agents. In a communication network setup, partial observability and/or communication constraints necessitate the learning of decentralized policies, which relies only on the local action-observation history of each agent. Decentralized policies also avoid the exponentially growing joint action-space with the number of agents, and therefore more practical and faster to converge in training. Fortunately, decentralized policies can be learned in a centralized fashion specially in a
simulated or controlled environments. The centralized training has access to hidden state information of other agents and removes inter-agent communication constraints. The paradigm of \textit{centralized training with decentralised execution} has already attracted attention in the RL community \cite{lowe2017MAACCC, kraemer2016multi,oliehoek2016concise}. 
% However, many challenges surrounding how to best exploit centralized training remain open.

Note that while the centralized training is offline training in a simulated environment but this is similar to the available networking and routing protocols as they are designed and optimized based on the modeled environments and network simulations. The protocols may be adaptive but the rules (even the adaptation process) are all designed offline. More importantly, the routing path parameters ($C$ and $H$ values) are regularly updated in real-time (during execution) based on the length and confidence of potential paths and the dynamic of the network. Although our DNN-based routing policy designed based on models and simulations (similar to any other rule-based human-crafted or machine learning-based routing policy), the policy decisions are based on online input features collected through the CQ+ path parameters ($C$ and $H$) propagation process through ACK messages and it considers all dynamics of the environment in real-time.

\subsubsection{Parameter Sharing}
In the context of centralized training but decentralized execution, a common strategy is to share the policy parameters between agents that are homogeneous \cite{gupta2017cooperative, terry2020parameter, chu2017parameter} (parameter sharing). Note that in the heterogeneous communication networks, we can always categorize nodes as homogeneous with some individual parameters and those can be also given as input to the policy with shared parameters. 

\begin{figure}[t]
\centering
\includegraphics[width=8.5cm]{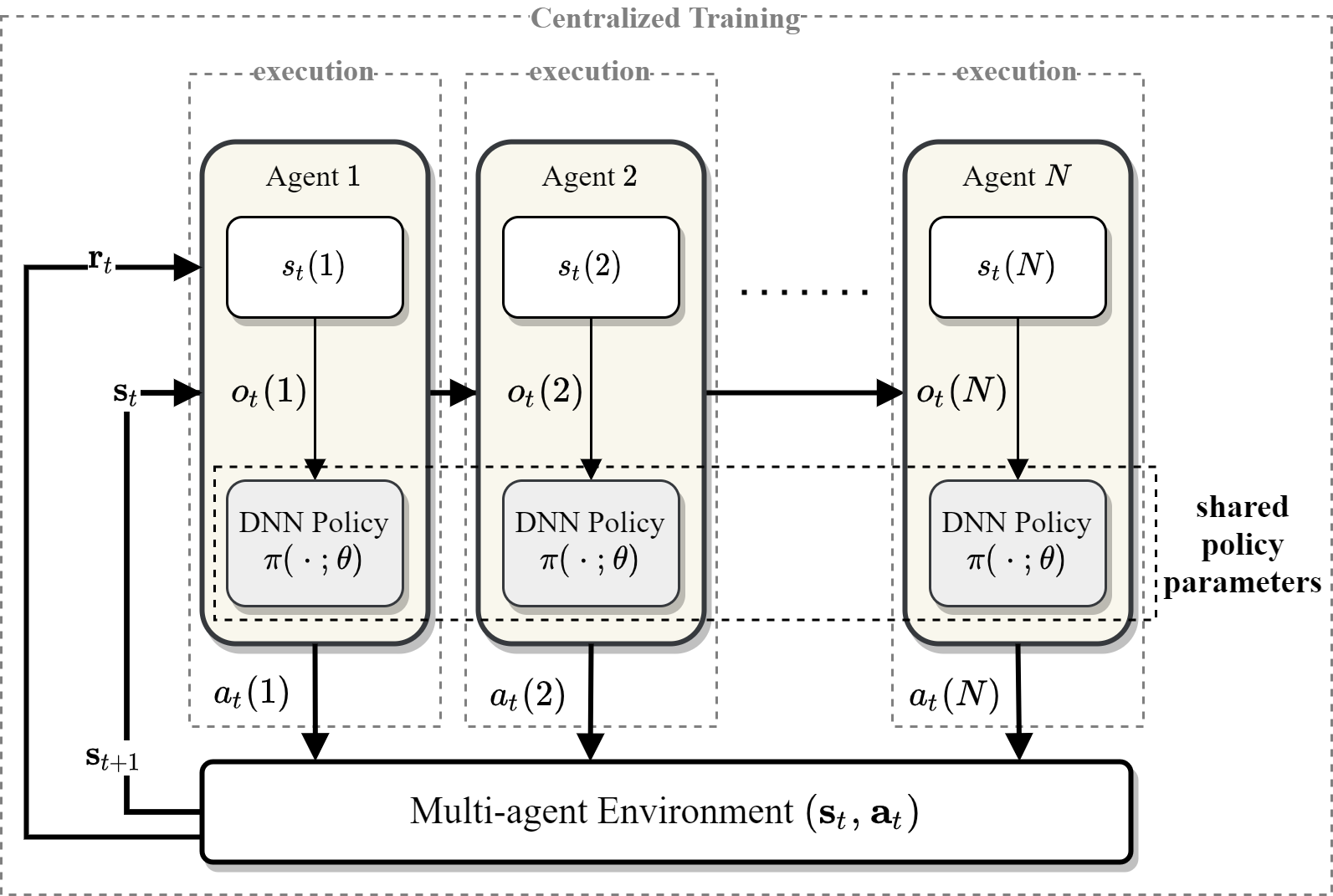}
\caption{Multi-agent network routing environment with shared policy parameters between agents. The centralized training and decentralized execution are also shown. Each agent $i$, uses the shared policy $\pi_\theta$ individually to find its own action $a_i(t)$ based on its own observations $o_i(t)$. The multi-agent environment operates based on the joint-actions decided and taken individually and transition to next state $\mathbf{s}_{t+1}$ and the rewards are pulled out based on that.}
\label{decentralized_exec_centralized_train}
\end{figure}

The multi-agent environment for the routing problem and our proposed framework are also summarized in Fig. \ref{decentralized_exec_centralized_train} where the centralized training, decentralized execution, and parameter sharing are illustrated.

\subsection{Our proposed DRL Framework and Approach}
We consider wireless communication networks with variable sizes (e.g. $5 \leq N \leq 50$). There can be multiple flows in the network with variable source/destination pairs. At any time, a node $i$ where $1 \leq i \leq N$ in the network may hold a packet(s) in its queue to be delivered to a destination node $d$. Note that node $i$ may or may not be the original source of that data packet. The network is highly dynamic in which the nodes are consistently moving at various random velocities and directions. The dynamic level of each node is denoted by $\bar{v}$ represents the average speed of that node. The dynamic level can be different depending on environment, scenario, and vehicle type of the node. We used reliable MANET mobility models such as Gauss-Markov model and/or random way-point for the nodes' mobility, which is discussed in Section \ref{section:mobility}. At each time $t$, each node $i$, picks a packet from its data queue (according to a certain packet scheduling algorithm). We use a simplified first-in-first-out (FIFO) packet scheduling but the routing decisions and policy design are independent of this scheduling method. Each node holds a table of network parameters represented by two matrices of $C$- and $H$-levels discussed in Section \ref{section:r2dn_protocol}. In other words, each node $i$ has two 2-dimensional $(N-1) \times (N-1)$ matrices holding $\mathbf{C}_i$ and $\mathbf{H}_i$ with entries of $c(i,j,d)$ and $h(i,j,d)$ where $1 \leq j,d \neq i \leq N$ respectively.
\begin{equation}
    \mathbf{C}_i = [c(i,j,d)] \quad \text{for all } 1 \leq j \neq i, d \neq i \leq N
\end{equation}
\begin{equation}
    \mathbf{H}_i = [h(i,j,d)] \quad \text{for all } 1 \leq j \neq i, d \neq i \leq N
\end{equation}
Note that the entries are associate to $N-1$ other nodes $j$ (set of all possible next-hops) and $N-1$ possible destination nodes $d$ (other than $i$). These matrices are initialized when an episode is started. Hence, the network parameter matrices $\mathbf{C}_i$ and $\mathbf{H}_i$ are often sparse as each node may have a limited number of neighbors and packet destinations and therefore most of the entries of these matrices often remain as their initial reset values.

For the sake of simplicity of notations, the time length of data packets are considered fixed and it is used as our units of time. Each realization of the network (also called an \textit{episode} in the context of RL) has at least $T$ time-slots (1 packet duration time). Note that the packet duration time is assumed fixed but the data rate can still be different. An episode is ended when the length of the episode exceed a maximum traffic length, $T_{max}$, and all the nodes have empty data queues. There is no incoming traffic beyond $T_{max}$ and after than we only wait until all nodes being done with their routing decisions. The goodput rate is computed as the ratio of the number of delivered packets to the total incoming packets in the episode. We exclude duplicate delivery of packets at destination if it happens.

Each node $i$ has a packet routing policy $\pi$ which assigns a routing decision for each packet based on the current state of the node. Since we use parameter sharing when we perform training, we consider the same parameterized policy $\pi_\theta$ used at every node (one policy for all). This assumption helps with the scalability of our framework as the designed policy will be used for any network sizes and for every node and there is no need for re-train and re-design of the policy. When the policy function is defined as a DNN, the parameter vector $\theta$ represents the weights of the neural network.   

% This is indeed consistent with the original SRR policy as it offers the same policy for all, although it is a rule-based probabilistic decisions only (not DNN) (see (\ref{r2dn_policy})).

The main distinction of our work from similar DRL-based routing algorithms is that our solution is scalable in terms of \textit{(i)} communication network size \textit{(ii)} network dynamics and mobility levels and topology \textit{(iii)} different data flows (source and destination pairs). 
We train our DNN policy for a single network size, single data flow (single source and destination pair), and small range of dynamic levels but the performance is tested and verified for variable network sizes, multiple data flows, and larger range of dynamic levels. 

\subsubsection{Pre-processing of the policy input features}
To satisfy the scalability requirement, the proposed DNN policy needs to have a network architecture to accommodate variable number of agents. The routing decisions are generally dependent on the network parameters (i.e. $C$- and $H$-values) of the neighbors. For example, if a subset of agents have zero $C$-levels then there is no need to play them in our routing decisions. Hence, we perform a pre-processing of the network parameters available (state characteristic) to have fixed number of inputs fed into our DNN policy. We select the best $K$ neighbors out of total $N-1$ possible neighbors in the network for each node and use their network parameters in the current state definition of that specific node. For the sake of simplicity, we drop the current node $i$, and destination $d$ from $C$- and $H$-levels and use the next-hop as index (i.e. $c_t(i,j) := c_t(i,j,d) $). We also order the next-hop indices according to ascending order of $h_j(1-c_j)$. Now, in our problem formulation the best $K$ neighbor $C$ and $H$ values are represented by 
\begin{equation}
    \begin{split}
\mathbf{c}_t(i) &= \left[c_t(i,i_1), c_t(i,i_2), \ldots, c_t(i,i_K)\right] \\
\mathbf{h}_t(i) &= \left[h_t(i,i_1), h_t(i,i_2), \ldots, h_t(i,i_K)\right]
    \end{split}
\end{equation}
where the neighboring nodes $i_1,...,i_K$ of $i$ are ordered so that
\begin{equation}
\label{preprocess_CH_sorting}
    h_t(i,i_1)(1-c_t(i,i_1)) \leq \cdots \leq h_t(i,i_K)(1-c_t(i,i_K)).
\end{equation}
Note that we choose a fixed value of $K$ for the range of network sizes we have considered in this work ($5 \leq N \leq 30$). A single value of $K$ (e.g. $K = 4$) is shown to be enough in terms of scalability. Also, note that we select the best neighbors (or other nodes) based on the smallest multiplication of uncertainty and number of hops and it is not arbitrary $K$ neighbors (or other nodes). For the other nodes that are not selected either, we do not have any information updated from them or their confidence level is too low and/or their expected distance to the destination is very high. Ideally, we are interested in the smallest value $K$ that training DNN policy based on it scales well for larger networks and a wider range of network parameters.
\subsubsection{State/Observations}
 Since we use a fully connected neural network (FCNN) rather than recurrent units, we want to capture the temporal changes of the selected $C$ and $H$ levels in the input features (as observations) given to the FCNN policy. Therefore, we add the change between current observations and previous observations, and also the previous effective action taken by the agent to the observations at node $i$ that is fed into the DNN policy as input features. Including the previous observations for the input features helps to capture temporal change rate of network parameters. Alternatively, we can benefit from recurrent neural network architectures but we have found that training such networks are typically takes longer. The input features to the DNN policy at node $i$ are given by
\begin{equation}\label{input_features}
    \mathbf{o}_t(i) = \left[\mathbf{c}_t(i), \mathbf{h}_t(i), \Delta\mathbf{c}_t(i), \Delta\mathbf{h}_t(i), a_{t-1}(i), p_{t-1}(i)\right],
\end{equation}
where $\Delta\mathbf{c}_t(i) = \mathbf{c}_t(i) - \mathbf{c}_{t-1}(i)$, $\Delta\mathbf{h}_t(i) = \mathbf{h}_t(i) - \mathbf{h}_{t-1}(i)$, and $a_{t-1}(j)$ is the previous action of some node $j$ that the current packet is received from. The last indicator shows if the current packet is received as a result of another node's broadcast or unicast. 

\begin{table}
\begin{center}
 \begin{tabular}{l  c  c  c } 
 \hline 
 Routing Protocol & C/Q-values & Broadcast & MADRL \\ 
 \hline \hline
 CQ-routing \cite{kumar98} & $\checkmark$ & $\times$ & $\times$ \\ 
 \hline
SRR (CQ+) \cite{johnston18} & $\checkmark$ & $\checkmark$ & $\times$ \\
 \hline
  DeepCQ+ routing (this work) & $\checkmark$ & $\checkmark$ & $\checkmark$   \\
  \hline 
\end{tabular}
\end{center}
\caption{Comparison of robust routing protocols in dynamic networks}
\end{table}

\subsection{Reward Definition for DeepCQ+ routing}

\subsubsection{SRR (CQ+ routing) as a DRL-based Policy}
In this section, we reformulate the RL problem so that its optimal policy results in the same policy as the CQ+ routing policy. This particularly helps in defining a proper rewarding system for our CQ+ routing policy improvements and extensions. 

First, we consider a stochastic routing policy $\pi(a|s_t)$ where the action's probabilities are only dictated by the policy. This is indeed similar to the CQ+ routing routing policy where it computes a probability of broadcast. We give the following rewards to the unicast action ($a = 0$) and the broadcast action ($a = 1$) as
        \begin{equation}
        \label{SRR_rewards}
            r_t(\mathbf{s}_t, a_t) = 
            \begin{cases}
             1-c_t(i,i_1)\tilde{\epsilon} & a_t = 1(\text{broadcast}) \\
             c_t(i,i_1)\tilde{\epsilon} & a_t = 0(\text{unicast})
            \end{cases}
            \quad \text{(Reward 1)}
        \end{equation}
where $\tilde{\epsilon} = 1- \epsilon$. Note that $c_t(i,i_1)$ is closely tracks the success/freshness probability of the path via the next-hop. Then, in order to maximize the total expected (discounted) reward as
        \begin{equation}
        \label{SRR_exp_reward}
        \begin{split}
        R = \mathbb{E}_{s_t\sim p^\pi, a_t \sim \pi}&\left[\sum_{t=0}^T \gamma^tr_t\right] =  \\
        \mathbb{E}_{s_t\sim p^\pi, a_t}&\left[\sum_{t=0}^T \gamma^t\left( \pi(a_t=0|\mathbf{s}_t)c_t(i,i_1)\tilde{\epsilon} \right.\right. \\
        &+\pi(a_t=1|\mathbf{s}_t)(1-c_t(i,i_1)\tilde{\epsilon})) \Bigg]
        \end{split}
        \end{equation}
        
Now, with zero horizon (i.e. $T = 0$), maximizing the expected reward given by (\ref{SRR_exp_reward}) leads to a policy with probability of broadcast is 
    \begin{equation}\label{hard_SRR}
    \pi(\mathbf{s}_t,a_t = 1) = P_\text{BC} = \begin{cases}
     1 & c_t(i,i_1)\tilde{\epsilon} < 1/2 \\
     0 & c_t(i,i_1)\tilde{\epsilon} > 1/2.
    \end{cases}
    \end{equation}
This is indeed the deterministic version of the CQ+ routing routing policy previously discussed as in (\ref{r2dn_policy}).

An immediate improvement of the CQ+ routing policy using DRL can be pursued as maximization of the expected return, $R$ as in (\ref{SRR_exp_reward}) with $\gamma > 0$ and some large horizon $T$, where the immediate reward $r_t(\mathbf{s}_t, a_t)$ is given by \ref{SRR_rewards}. We can use efficient RL algorithms such as PPO to find a DNN policy. We refer to this DNN-based version of CQ+ routing as DeepCQ+ routing with reward type 1. Note that different reward designs will give different performance objectives. 

The CQ+ routing policy aims to choose the next-hop that minimizes uncertainties (probability of failure) in dynamic networks. This is done by accounting for the probability of failure for the path with the lowest link uncertainties and number of hops. Although the CQ+ routing policy indirectly optimizes the network overhead, it is not accounted for directly in its optimization. We define the overhead in the routing of the dynamic networks as the ratio of the total number of transmissions in the communication network, denoted by $N_\text{TX}$, to the total number of packets delivered, denoted by $N_D$ for a specific packet data rate and a window of time. The \textit{normalized overhead} by the network size, $N$, is given by 
    \begin{equation}
        \mathsf{OH} = \frac{1}{N} \cdot \frac{N_\text{TX}}{N_\text{D}}.
    \end{equation}

To show the effectiveness of the DRL approach on the CQ+ routing policy optimization, the DeepCQ+ routing objective is to minimize overhead while keeps the goodput rate $\rho$ at the same level (or higher) than the goodput rate provided by CQ+ routing, i.e. $\rho_0$ (for the same input data flows and the same horizon). This can be formulated as 
    \begin{equation}
    \label{overhead_minimization}
        \min\limits_{\theta}  \frac{N_\text{TX}}{N_D} \quad  
        \text{ such that } \rho \geq \rho_0.
    \end{equation}
Since the target is the efficiency of our routing policy, we can simply assume $\rho = \rho_0$, and consequently, $N_D = \rho_0 T$ for some time horizon $T$ (for some specific input data rate). With this approach, we maintain the goodput rate while we achieve lower overhead and it is fair to compare with previous robust routing policies. Therefore, the optimization problem in (\ref{overhead_minimization}) can be simplified as
\begin{equation}
\label{overhead_minimization_2}
    \min\limits_{\theta}  N_\text{TX} \quad \text{ such that } N_D = \rho_0T
\end{equation}
From our experiments, using the rewarding system in (\ref{SRR_rewards}) gives us the same goodput rate as $\rho_0$ while keeping the overhead lower than CQ+ routing. Hence, we can consider the overhead minimization by reformulating our rewarding system to accommodate (\ref{overhead_minimization_2}) as
% \begin{equation}
% r_i(t) = w_1\mathds{1}_{\mathcal{D}}-w_2D - w_3\mathds{1}_{\mathcal{Z}} -w_4\frac{N_\text{ack}}{N} \quad \text{(Reward 2)}
% \end{equation}
\begin{equation}
r_i(t) = w_1\mathds{1}_{\mathcal{D}}-w_2\mathds{1}_{\mathcal{Z}} -w_3\frac{N_\text{ack}}{N} \quad \text{(Reward 2)}
\end{equation}
where $\mathds{1}_{\mathcal{D}}$ is the reward for packet delivery. If a packet is delivered successfully to the destination and node $i$ has contributed to that delivery then it will be rewarded. %Also, $D$ is the end-to-end delay when the packet is delivered. Note that we do not reward for duplicate delivery of a packet. 
Also, $\mathds{1}_{\mathcal{Z}}$ is a reward (penalty) indicator which is enabled when we have not received any ACKs from our transmissions. This term is not directly related to the optimization problem (\ref{overhead_minimization_2}) but it prevents the system to learn unicasting all the packets initially, which may happen if the initial reward is negative and the agent wants to end the episode to prevent increased penalties. The next term in the reward, i.e. $N_\text{ack}/N$, is the normalized number of ACKs received as a result of the action taken and therefore closely represents the number of copies of a packet at other nodes. A naive approach can be to penalize for every transmission directly. However, the cause of transmissions at each node is not the actions taken at that specific node, but it is the transmissions of the packets from other nodes to that specific node. Hence, $N_\text{ack}$ received at each node is used to estimate the added number of transmissions dictated to the network as a result of each node's action. The last reward component remains intact from the reward type 1 as it was already outperforming CQ+ routing in terms of overhead while achieving the same goodput (see Fig. \ref{fig:goodput_rate} and Fig. \ref{fig:overhead_rate}). The weights of the reward components, (i.e. $w_1, w_2, w_3$), have been tuned according to the overhead minimization problem (\ref{overhead_minimization_2}). The proposed DeepCQ+ routing technique with modified (extended) reward definition is referred to as DeepCQ+ routing with reward type 2. The DeepCQ+ routing algorithms are summarized in Algorithm \ref{DeepR2DN-algorithm}. 

\begin{algorithm}
\SetAlgoLined
\caption{The proposed DeepCQ+ routing (with reward type 1 and 2)}
\label{DeepR2DN-algorithm}
\small
{Receive incoming packet at node $i$:} \newline
\eIf{Packet is ACK}{Update $c$ and $h$ using (\ref{update_H}) and (\ref{update_C})}
    {\If{packet traversed a loop}{Drop packet, do not return ACK}
     \If{packet is already in queue}{
        Drop packet \\
        Find best next-hop $i_1$ from (\ref{best_next_hop})) \\
        Compute $c_\text{ack}$ and $h_\text{ack}$ from (\ref{c_ack}) and (\ref{h_ack}) using $j^\star$ \\
        Return ACK}
          \eIf{packet is not duplicate}
        {Add packet to the queue}
     {Do not add packet to the queue}}
    \If{ACK Packet is not received}{Do not update $c$ and $h$, (i.e. no packet is received)}
    \If{Queue is not empty}{Pick up packet from queue of node $i$ (at current time $t$) \\
    Pre-process best ($K=4$) neighbors using (\ref{preprocess_CH_sorting}) \\
    Form the input to the DNN-based routing policy \\
    $\mathbf{o}_t(i) = [\mathbf{c}_t(i), \mathbf{h}_t(i), \mathbf{c}_{t-1}(i), \mathbf{h}_{t-1}(i), a_{t-1}(i)]$ \\
    \mbox{\textsc{Routing Decision Policy}} \\
    DeepCQ+ routing (reward 1 or 2): $\theta= \theta_\text{DeepCQ+ routing-rew-1 or -2}^\star$
    Choose $\begin{cases}
     \text{Broadcast} & \text{with probability } \pi_\theta(a = 1|\mathbf{o}_t(i); \theta) \\
     \text{Unicast} & \text{with probability } \pi_\theta(a = 0|\mathbf{o}_t(i); \theta)
     \end{cases}$ \\
     \If{Decision is Broadcast}{Forward packet to all}
     \If{Decision is Unicast}{Forward packet to $i_1$}}
\end{algorithm}

\section{Experiments and Numerical Results}
\subsection{Environment Modelling and Training Platforms}
To model the CQ+ routing environment and for rapid development of the design, algorithms, and testing, we have developed a CQ+ routing network simulator and constructed an RL CQ+ routing environment to train and test our approach and the CQ+ routing baselines. Our environment platform is built in Python and it has all CQ+ routing protocol features including generation of random dynamic networks with data flows, different number of nodes, and configurable network setup parameters (dynamic levels, range, node locations, data rates, source and destinations, data queues and backlogs, link quality computation, etc.). The CQ+ routing protocol contains the tracking, distributing $C$- and $H$-values, duplicate packet checking, etc. It also extracts performance evaluation information. The environment is interfaced with the \textit{Ray}, which is a powerful distributed computing platform for the machine learning \cite{moritz2018ray} and RLlib library \cite{liang2018rllib} which provides scalable software primitives for the RL algorithms.
\begin{figure*}[t]
\includegraphics[width=\textwidth]{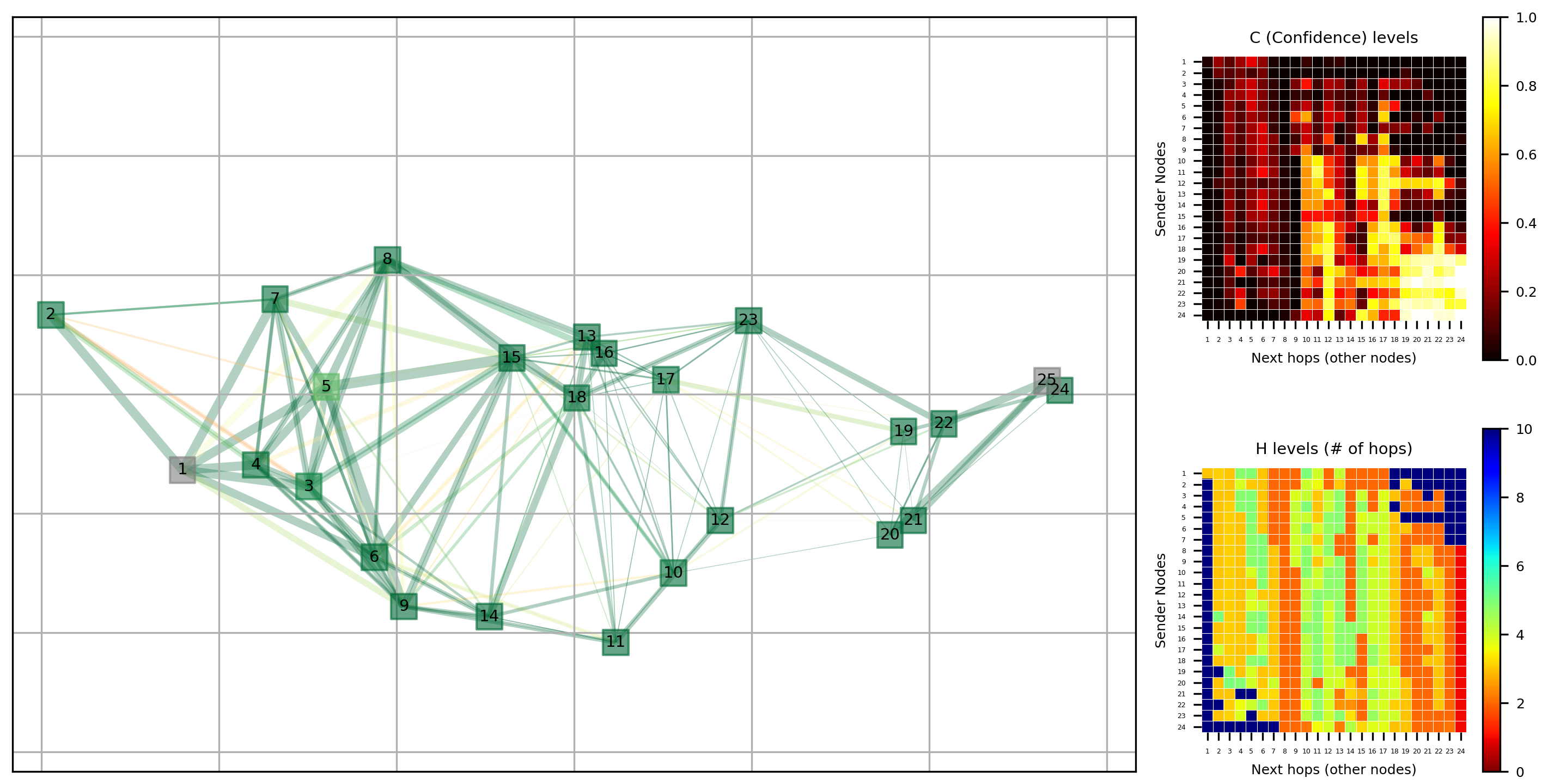}
  \caption{An example of our developed CQ+ routing environment (simulator) for a 25-node dynamic network.}
  \label{R2DN_simulator}
\end{figure*}
A snapshot of the environment is given in Fig. \ref{R2DN_simulator}, where the source and destination are in colored as gray. The color of the nodes shows the queue length or the congestion level of the nodes. The color of the links is associated with their quality with green links represents high quality and low packet error rates. The thickness of the links shows the traffic level of the link. The heatmap images of the $C$- and $H$-levels are monitored to show the state of the network visually. The heatmap images of $C_i(d, j)$ are 2-D vectors as they consider single destination node (i.e. node $d = 25$). Each row is associated with the current node $i$ and each column corresponds to the other nodes $j$ of each current node $i$. For example, the $C$-levels for the other nodes that are closer to the destination ($j \geq 20$) are relatively high as there is high confidence of the path going to the destinations through them.

\subsubsection{Benchmark Topology}
A benchmark topology was considered based on the adaptive routing work \cite{danilov12} and the CQ+ routing \cite{johnston18}. In this topology scenario, there are $N$ ($N \geq 30$) nodes considered in the network which are randomly located in an area. The nodes between the source and destination nodes are moving in random directions with variable speeds. The closer they are to the source or destination, the slower their speed is, reflecting the heterogeneous tactical network environments. An example that is considered in the training and testing in the area of 800 (m) by 300 (m) with the wireless range of the nodes is 150 (m). The packet error rates drop rapidly when the distance between nodes exceeds that range. We assume no interference between transmissions to focus on the routing layer. Source and destination nodes exhibit a slow mobility pattern, similar to the nodes in the outermost regions. The data flow between the source and destination nodes can have various data rates (e.g. 20 packets per second at up to 1000 bytes each; each flow generates up to 160Kbps payload traffic). In different experiments, we use different routing policies starting from the generic CQ+ routing policy \cite{johnston18}. The same network topology will be used to compare other routing algorithms such as DeepCQ+ routing with reward types 1 and 2. For both testing and training (to monitor the policy training progress), we monitor the resulting performance metrics such as broadcast rate (the percentage of the broadcast actions in the network), the goodput rate. Goodput is the rate of the delivery of intended data packets that are successfully received at the destination. We exclude any duplicate packets from counting as delivered packets, and in the RL context, we do not count them towards any sort of reward. More importantly, we consider the total overhead as a metric that quantifies the efficiency of our decisions and it is the ratio between the total number of data packets delivered to the total transmissions that have been made. Other network metrics are also recorded, such as the average number of hops between source and destination. Factors such as overhead and the number of hops are normalized by the size of the network so they apply to networks of different scales. 
% \begin{figure}[b]
% \includegraphics[width=8.5]{Figures/benchmark_topo_1.png}
%   \caption{A benchmark topology where a large network of nodes are connected and multiple data flows are injected from a source to a destinations selected from two ends of the network. The network is dynamic and the routers in the central regions are moving faster.}
%   \label{R2DN_topology}
% \end{figure}
\subsubsection{Mobility Model}\label{section:mobility}
In the mobile ad-hoc networks (MANET), there have been various mobility models proposed and discussed recently and particularly in their impact on the network routing protocols \cite{gupta2013performance, timcenko2009manet, Ariyakhajorn2006,divecha2007impact,bai2004survey}. The MANET mobility models represent the moving behavior of each mobile node in the MANET. MANET mobility models are particularly important from our training and testing perspective. These models need to be realistic for the dynamic tactical wireless networks, while still need to have enough randomization to cover corner cases and extreme scenarios in our training and properly reflect the performance impact of the network dynamics, scheduling, and resource allocation protocols. The random waypoint mobility model is often used in the simulation study of MANET, despite some unrealistic movement behaviors, such as exhibiting sudden stops and sharp turns \cite{bettstetter2004stochastic}. The Gauss-Markov mobility model \cite{liang1999predictive} has been shown to solve both of these problems \cite{Ariyakhajorn2006}. Therefore, we have used this model in our training and testing process, and implemented in our network simulator platform. To be complete, we have also tested the trained policies using the Gauss-Markov model on the networks that follow random way-point model and observed similar performance and behavior. Compared to the random way-point model, the Gauss-Markov mobility model has improved modeling performance at higher dynamics (e.g. as fast as fast automobiles) while it holds the same performance as a random way-point at human running speeds \cite{Ariyakhajorn2006}. For the training of the MANET, the important factor is the sensitivity of the throughput (or goodput) and the end-to-end delay to the different levels of the randomness settings, and the Gauss-Markov model shows no effect on the accuracy of these metrics \cite{Ariyakhajorn2006}.

In the Gauss-Markov model, the velocity of mobile node is assumed to be correlated over time and modeled as a Gauss-Markov
stochastic process. In a two-dimensional simulation and emulation field (as in this study), the value of speed and direction at the $n$th time instance is calculated on the basis of the value of speed and direction at the $n-1$th time instance and a random variable using the following equations: 
\begin{equation}
    v^{(n)} = \mu v^{(n-1)} + (1-\mu)\bar{v} + \sqrt{(1-\alpha^2)\tilde{v}^{(n-1)}}
\end{equation}
\begin{equation}
    \phi^{(n)} = \mu \phi^{(n-1)} + (1-\mu)\bar{\phi} + \sqrt{(1-\mu^2)\tilde{\phi}^{(n-1)}}
\end{equation}
where $v(n)$ and $\phi(n)$ are the new speed and direction of the node at time interval $n$; $0 \leq \mu \leq 1$ is the tuning parameter for the randomness (and correlation to previous time instance); $\bar{v}$ and $\bar{\phi}$ are constants representing the mean value of speed (i.e. dynamic level) and direction as $n \rightarrow \infty$; $\tilde{v}^{(n-1)}$ and $\tilde{\phi}^{(n-1)}$ are random values from a Gaussian distribution to add randomness. In a two-dimensional simulation and emulation field (as our study), the Gauss-Markov model gives the next location based on the current location at time instance $n$ as 
\begin{equation}
    x^{(n)} = x^{(n-1)} + s^{(n)} \cos(\phi^{(n)})
\end{equation}
\begin{equation}
    y^{(n)} = y^{(n-1)} + s^{(n)} \sin(\phi^{(n)})
\end{equation}
where $\left(x^{(n)},y^{(n)}\right)$ and $\left(x^{(n-1)},y^{(n-1)}\right)$ are the $x$ and $y$ coordinates of the MANET node's location at the $n$th and $(n-1)$th time intervals, respectively. The mean angle will be adjusted when nodes reach the region edges to limit the movement within the region.

\begin{figure}[h]
\centering
\includegraphics[width=8.5cm]{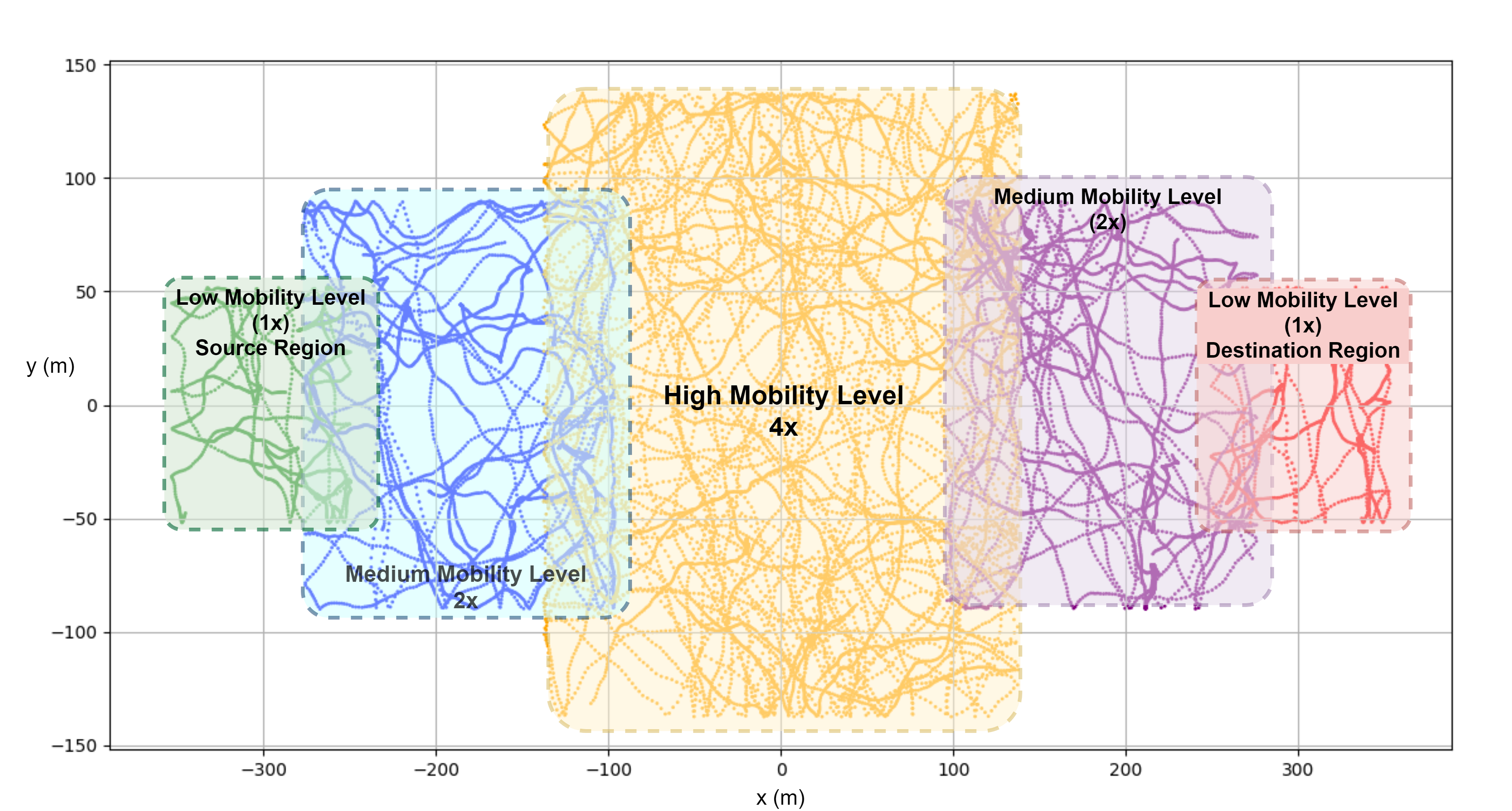}
\caption{Mobility regions and network topology. The movements of the nodes are shown for a network of size 30 according to the Gauss-Markov model}
\label{mobility}
\end{figure}

To more accurately reflect the benchmark topology used in the SRR (CQ+ routing) papers (discussed in the previous section), we have divided the environment into 5 groups symmetrically between the source and destination nodes. The source and destination regions are at the left and right corners of the area.   The closer the region is to the center, the faster the nodes move. Also, the regions are overlapping by \%10 to prevent too many network partitions. The central region has double the speed variance of that of the mid-left and mid-right regions. The source and destination regions have half the speed variance compared to the mid-right/left regions. The mobility of the MANET nodes is simulated and shown in Fig. \ref{mobility}.
\begin{figure}[h]
\centering
\includegraphics[width=8.5cm]{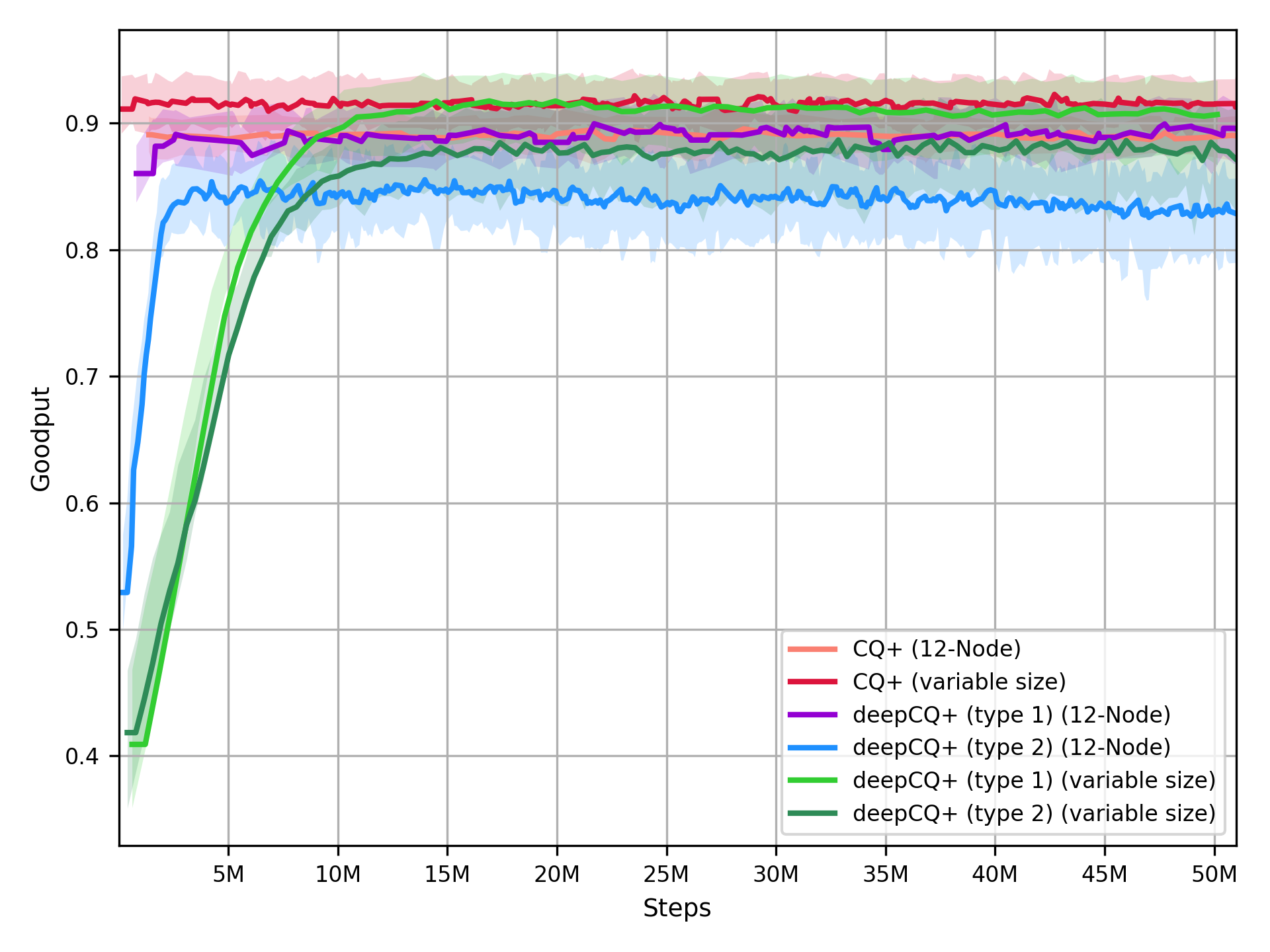}
\caption{Goodput rate (delivery rate) while training vs. number of steps.}
\label{fig:goodput_rate}
\end{figure}
\begin{figure}[h]
\centering
\includegraphics[width=8.5cm]{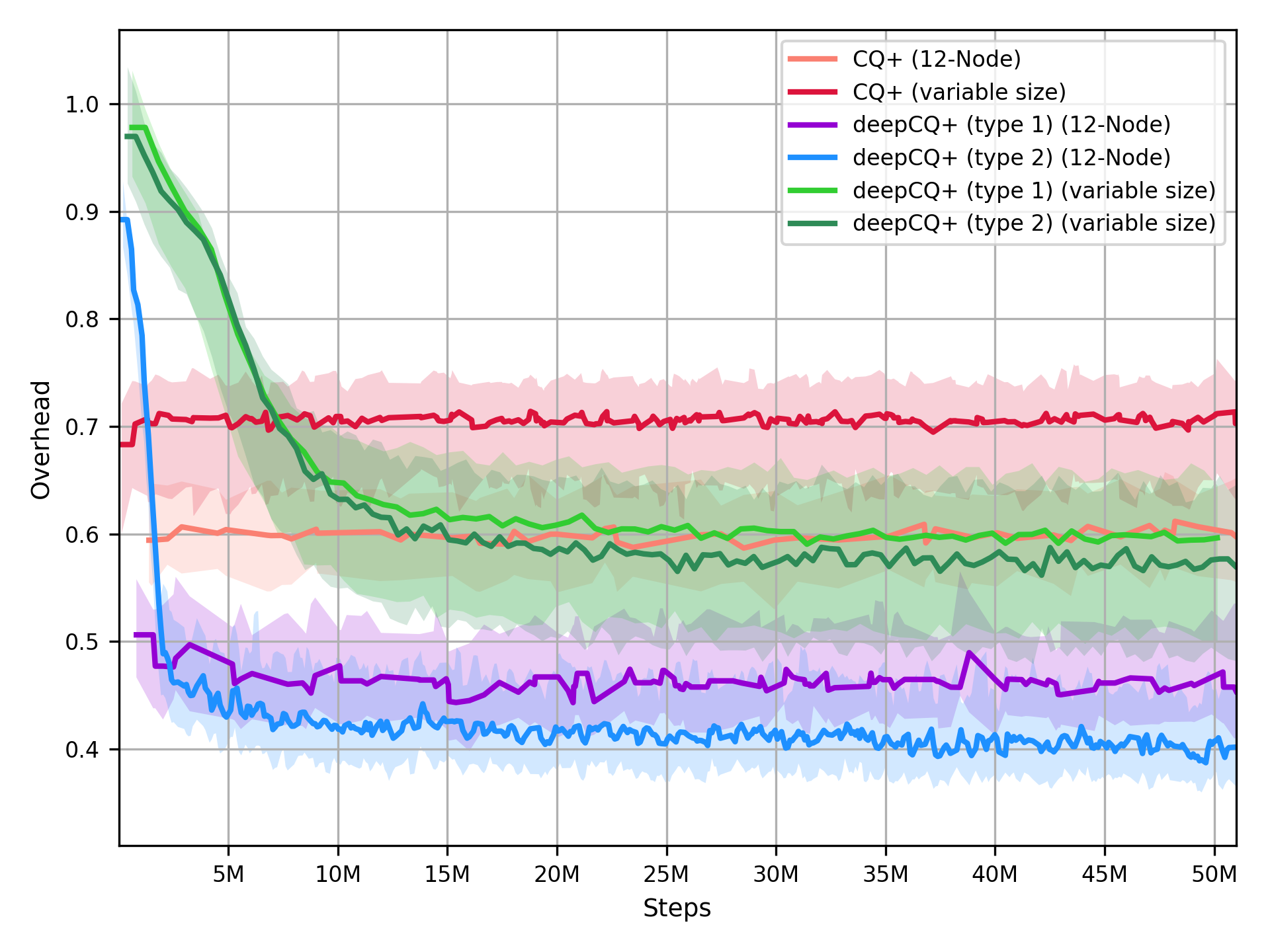}
\caption{Normalized overhead metrics during training vs. number of steps.}
\label{fig:overhead_rate}
\end{figure}
\begin{figure}[h]
\centering
\includegraphics[width=8.5cm]{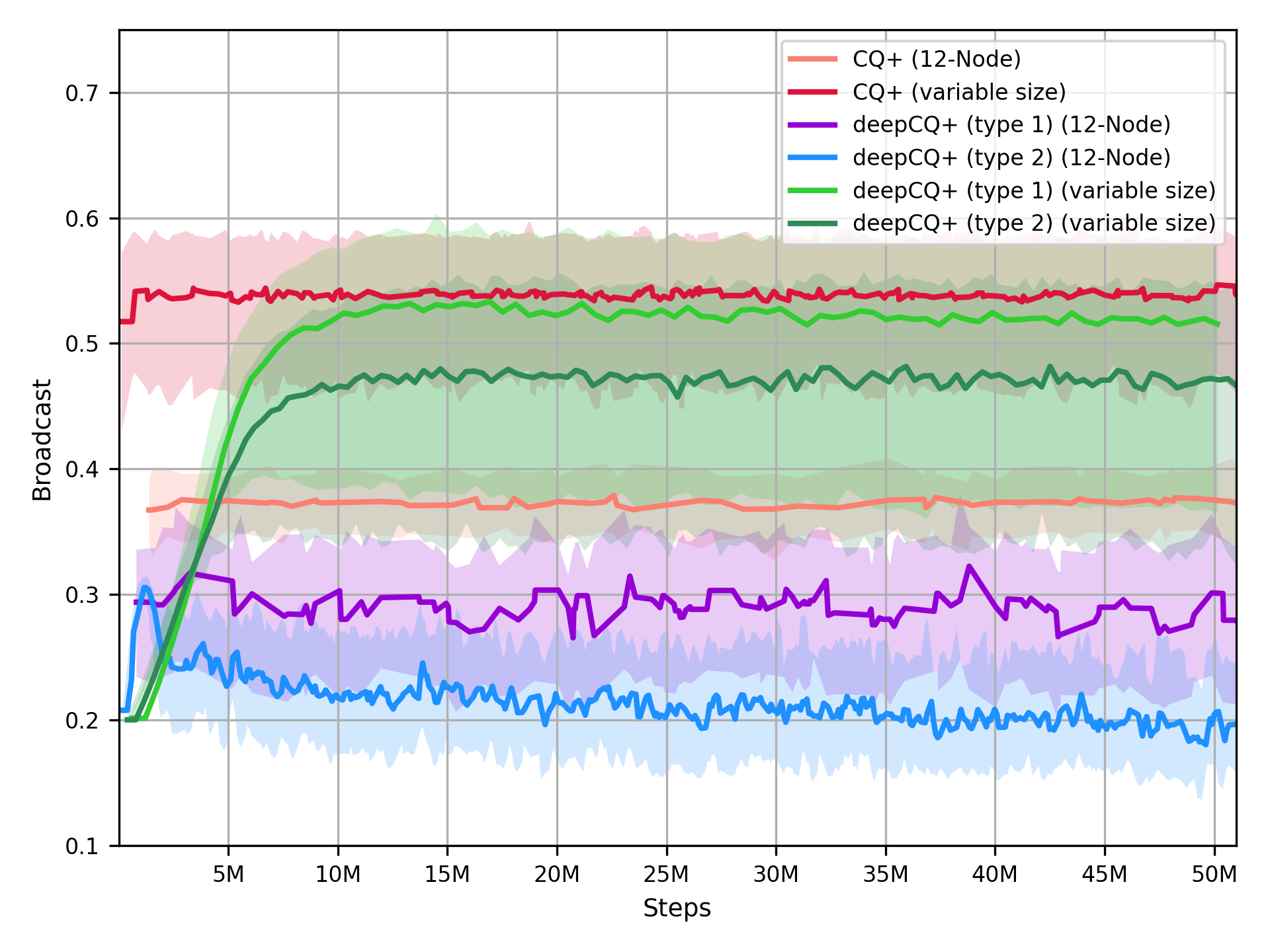}
\caption{Broadcast rate during training vs. number of steps.}
\label{fig:broadcast_rate}
\end{figure}
\subsection{Hyperparameter Tuning and Configuration Parameters}
In this section, we discuss the hyperparameter tuning process and list the parameters that obtained through tuning. We also itemize the configuration parameters used in the training process to generate the CQ+ routing environment (see Table \ref{tab:hyperparameters}).
\begin{table}[h]
\begin{center}
\caption{Numerical Results parameters and hyperparameters used in Fig. \ref{fig:deepVSCQ+}}
 \begin{tabular}{r  l} 
 \hline
 \textbf{Parameters} & \textbf{Value} \\ [0.5ex] 
 \hline\hline
 Network Sizes (Train) & 12 \\ 
 \hline
 Network Sizes (Test) & 5-50 \\ 
 \hline
 Learning Rate & 0.00005 \\
 \hline
 Discount Factor $\gamma$ & 0.99 \\
 \hline
 Episode Length & 3000 Packet Duration \\
 \hline
 Region Size Scale (Train) & 1 \\
 \hline
 Region Size Scale (Test) & 2 \\
 \hline
Dynamic Level Scale (Train) & 1 \\
 \hline
  Dynamic Level Scale (Test) & 5 \\
\hline 
 Policy DNN Size & FCNN(16, 16, 8, 8, 4) \\
 \hline
 Number of Data Flows (Train) & 1 \\
 \hline
 Number of Data Flows (Test) & 1,2,3, and 4 \\
 \hline \hline
\end{tabular}
\label{tab:hyperparameters}
\end{center}
\end{table}
\subsection{Numerical Results}
In this section, we discuss the results obtained during training and test of our DeepCQ+ framework versus CQ+. During training we use limited range of network settings, while we test on wider range of parameters as summarized in Table \ref{tab:hyperparameters} where we also listed the hyperparameters for the training of the DeepCQ+ policies. Note that to discuss the scalability, we also trained over wide range number of nodes (variable network size and up to 30) to show that training over wider range does not improve performance much and show that our MADRL approach is not over-fit. 
We have collected the performance metrics during training for over 50 million steps (over 15000 episodes) and shown in Fig. \ref{fig:goodput_rate}, \ref{fig:broadcast_rate}, and \ref{fig:overhead_rate}. The moving average curves are plotted along with the shaded variations of the performance metrics across training steps.

The results in Fig. \ref{fig:deepVSCQ+} confirms that in general, the DeepCQ+ routing with reward type 2 outperforms the generic (non-DRL based) CQ+ routing technique. Note that the test is performed over network sizes of 5 to 30 while the training is only performed on a single network size (e.g. $N = 12$ in these results). Note that one can choose (slightly) different value for the training network size but the scalability conclusion still holds.). This is evidence that the DeepCQ+ routing is scalable for different network sizes and dynamics once. DeepCQ+ performs as good as CQ+ routing, however DeepCQ+ with reward type 2 have lower overhead and lower broadcast rate. Note that the end-to-end delay component is not considered in either of reward types 1 and 2 and we just present it to show similar behavior with respect to delay. Note that the main message is the flexibility of our DRL framework to adjust the reward function (weights of various components including end-to-end delay) to optimize a policy for a certain trade space between goodput rate, overhead rate, and end-to-end delay. This was not available in CQ+ routing as it only statistically selects broadcast and unicast. The focus of the results in Fig. \ref{fig:deepVSCQ+} is to maintain goodput rate as high as CQ+ but improve the overhead. The normalized overhead is further divided by the network size to be fair in comparison across different network sizes and make the figure more readable. Our results show that the normalized overhead is at least 15\% lower in DeepCQ+ routing with reward type 2 compared with the non-DRL-based CQ+ routing while the goodput rate is about the same. Lower normalized overhead values are indicative of more efficient policies. Note that this is only an example of our DRL framework routing policy design to show achieving the target objective (e.g. minimize normalized overhead while maintaining goodput rate) and still being scalable (can be trained for certain network sizes and dynamics but performs satisfactorily for the rest of network parameters).
Note that the testing and training network settings are different as for every episode we use random locations, dynamic levels, network area size, and data flows and that is why there are variations during training and testing in the Figures. 

\begin{figure*}[h]
\centering
\includegraphics[width=\textwidth]{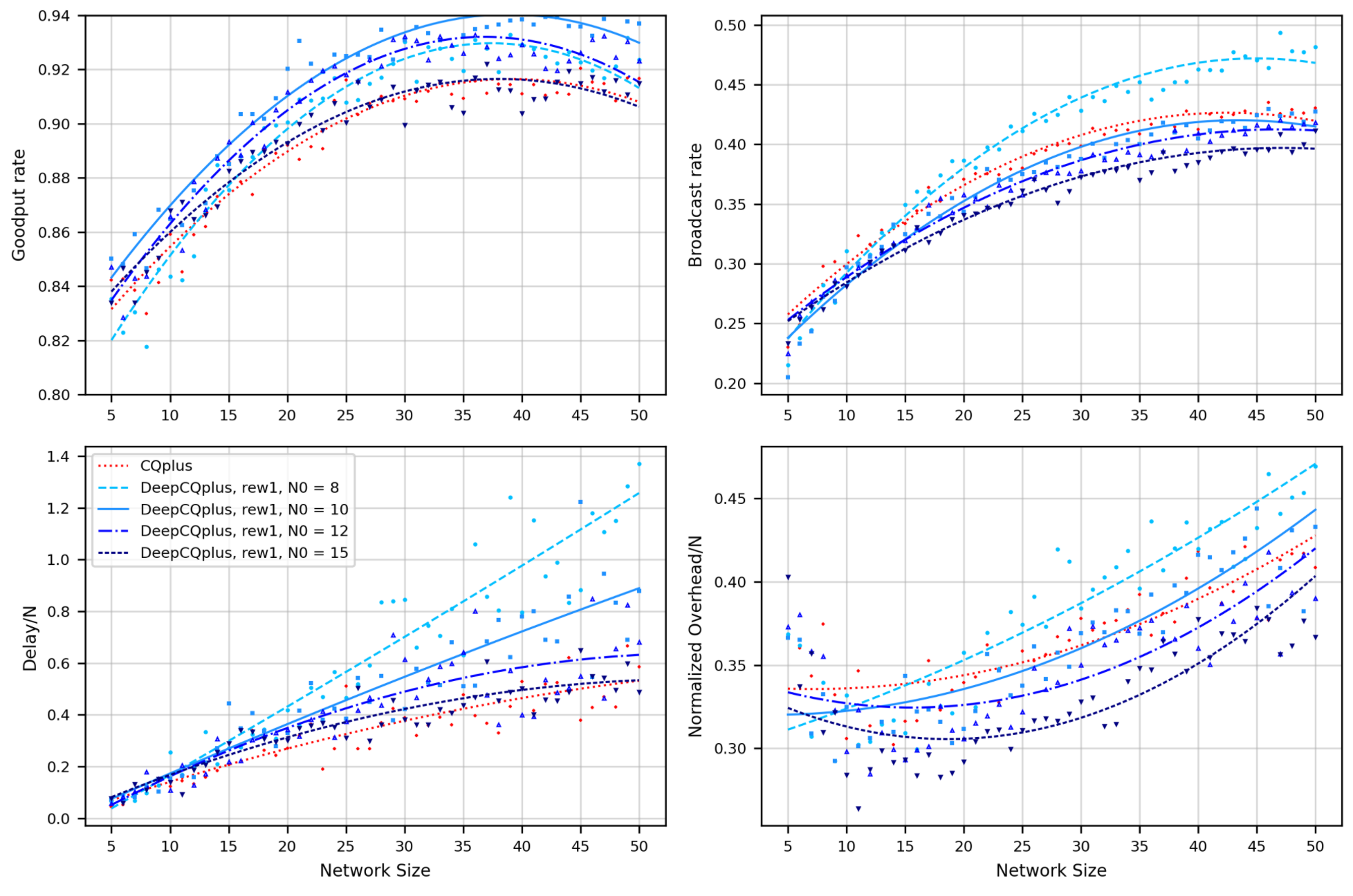}
\caption{Comparison of the results of the DeepCQ+ routing with reward type 1 when trained on various network sizes to show that training on 12-node gains most of the performance improvements of DeepCQ+. Nevertheless, this is to show the process to select 12 nodes but one may choose a different network size for training based on other network parameters. The test is averaged for 100 episodes for each number of nodes for the policy that is trained on 8, 10, 12, 15 nodes.}
\label{fig:deepVSCQ+}
\end{figure*}

\begin{figure*}[h]
\centering
\includegraphics[width=\textwidth]{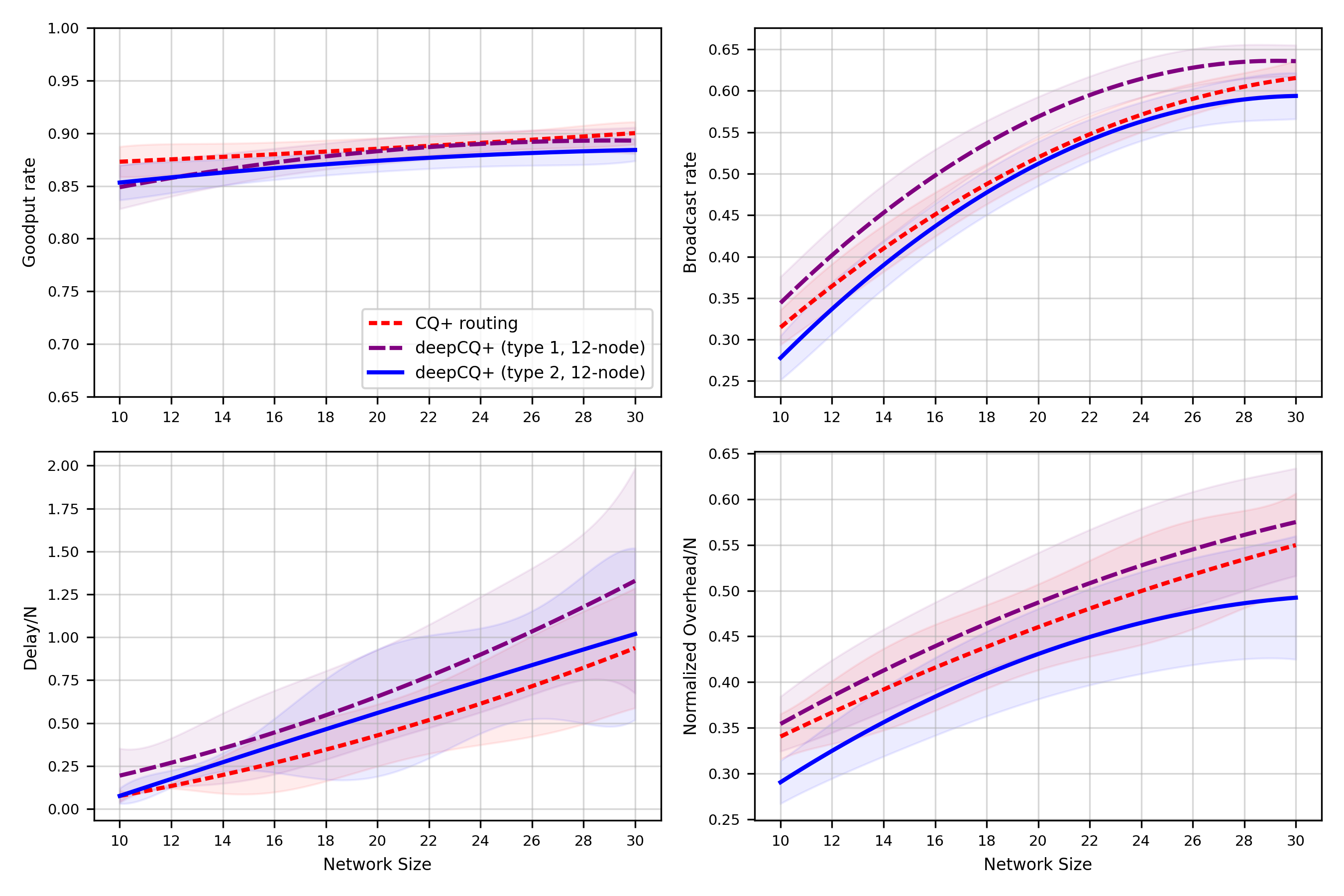}
\caption{Comparison of the results of the DeepCQ+ routing with reward type 1 and reward type 2 trained for a 12-node network only versus CQ+ routing; The results are tested across various network sizes from 10 to 30. Although the DeepCQ+ routing PPO policy is trained on 12-node networks, it scales perfectly for various network sizes. The DeepCQ+ routing with reward type 2 achieves significantly lower normalized overhead (overhead rate divided by the goodput rate divided by the network size). }
\label{fig:deepVSCQ+}
\end{figure*}
\begin{figure*}[h]
\centering
\includegraphics[width=\textwidth]{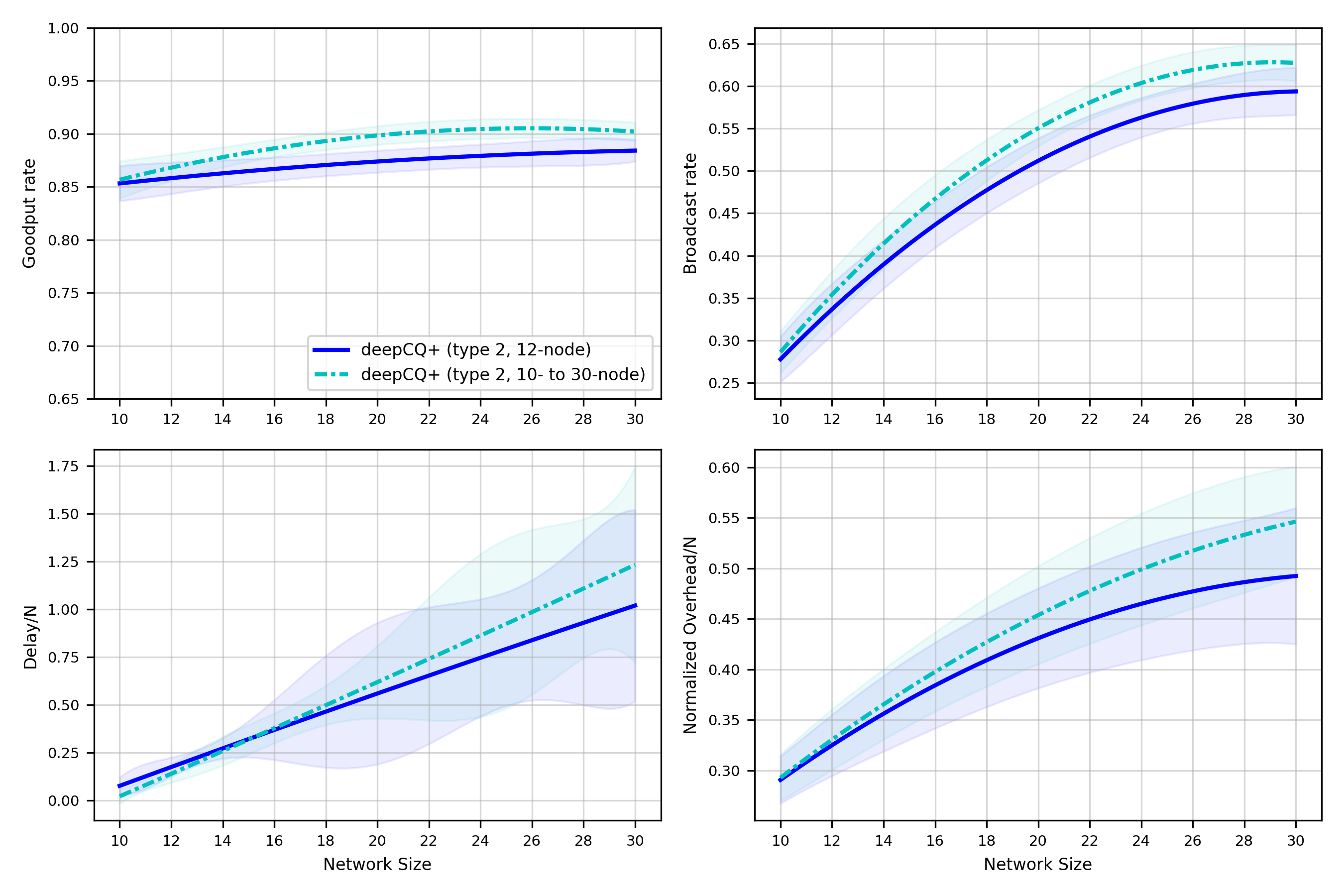}
\caption{Scalability of the DeepCQ+ routing with reward type 2 trained on 12-node networks is shown when compared with the policy trained over all 10- to 30-node networks (variable network sizes). The policy trained on 12-node shows as good as or even better in terms of normalized overhead on network sizes of 10 to 30.}
\label{fig:DeepCQ+12nodeScalability}
\end{figure*}
Our MADRL framework for the network routing policy design also enables us to train over variable network sizes and dynamics if it is required. Fig. \ref{fig:DeepCQ+12nodeScalability} shows the performance of DeepCQ+ with reward type 2 trained on 12-node networks and the same policy trained over variable network sizes 10 to 30. The results show that the training over 12-node networks scales perfectly for the entire network size domain and although it is possible to train over 10 to 30-node networks, there is not much gain in doing so if any. 

\section{Conclusions and Future Directions}
In this paper, we have demonstrated a successful and practical hybrid approach of MADRL and CQ+ routing techniques that is feasible to design a robust, reliable, efficient, and scalable policy for dynamic wireless communication networks, including many scenarios that the algorithm was \emph{not} trained for. Our MADRL framework, combined with the explainable CQ+ structure, is specially designed for scalability and enables us to train and test the routing policies for variable network sizes, data rates, and mobility dynamics with persistently high performance across scenarios that were not trained for. Our DNN-based robust routing policy for dynamic networks, DeepCQ+ routing, is based on the CQ-routing but also monitors network statistics to improve broadcast/unicast decisions. It is shown that DeepCQ+ routing is much more efficient than traditional CQ+ routing techniques, significantly decreasing normalized overhead (number of transmissions per number of successfully delivered packets). Moreover, the policy is scalable and uses parameter sharing for all nodes during the training, which allows it to reuse the same trained policy for scenarios with various mobility dynamics, data rates, and network sizes. It is noted that the sharing of parameters for all the nodes is not required during execution.

In our future works, we plan to expand the action space of the DeepCQ+ routing to include next-hop selection for the unicast mode. Other interesting extensions include further extending DeepCQ+ routing to accommodate heterogeneous wireless networks where a node may have multiple radio interfaces, extending the ACK-based information sharing to include additional context, and accommodating different performance metrics such as end-to-end delay minimization, overhead minimization, as goodput rate maximization. Another extension of the hybrid DeepCQ+ routing paradigm is to continue maintaining scalable and robust routing policies while being tuned to prioritize and balance network metrics to best meet the needs of virtually any MANET environment, including heterogeneous MANETs. 
\section{Acknowledgement}
Research reported in this publication was supported in part by Office of the Naval Research under the contract
N00014-19-C-1037. The content is solely the responsibility of the authors and does not necessarily represent the official views of the Office of Naval Research. The authors would like to thank Dr. Santanu Das (ONR Program Manager) for his support and encouragement. Also, we would like to thank the reviewers for their valuable comments to improve the quality of this paper.
% \subsection{CQ+ routing with next-hop selection}
% \subsection{R2DN with CQT(traffic)-routing}
% \subsection{Extend DeepR2DN to heterogeneous networks}
% \subsection{Modify exponential moving average of $C$ and $H$-values}
% \subsection{End-to-end delay minimization extension}

\medskip

\small
\bibliographystyle{IEEEtran}
\bibliography{DeepR2DN.bib}

\end{document}